\documentclass[10pt]{article}
\usepackage[colorlinks=true,urlcolor=blue,linkcolor=blue]{hyperref}

\begin{document}

\title{{\Large \textbf{Uniqueness of Petrov type D spatially inhomogeneous
irrotational silent models}}}
\author{Pantelis S Apostolopoulos and Jaume Carot \\
%EndAName
\\
{\small Departament de F\'isica, Universitat de les Illes Balears,}\\
{\small Cra. Valldemossa Km 7.5, E-07122 Palma de Mallorca, Spain}\\
{\small E-mails: pantelis.apost@uib.es; jcarot@uib.es}}
\maketitle

\begin{abstract}
The consistency of the constraint with the evolution equations for spatially
inhomogeneous and irrotational silent (SIIS) models of Petrov type I,
demands that the former are preserved along the timelike congruence
represented by the velocity of the dust fluid, leading to \emph{new}
non-trivial constraints. This fact has been used to conjecture that the
resulting models correspond to the spatially homogeneous (SH) models of
Bianchi type I, at least for the case where the cosmological constant
vanish. By exploiting the full set of the constraint equations as expressed
in the 1+3 covariant formalism and using elements from the theory of the
spacelike congruences, we provide a direct and simple proof of this
conjecture for vacuum and dust fluid models, which shows that the Szekeres
family of solutions represents the most general class of SIIS models. The
suggested procedure also shows that, the uniqueness of the SIIS of the
Petrov type D is not, in general, affected by the presence of a non-zero
pressure fluid. Therefore, in order to allow a broader class of Petrov type
I solutions apart from the SH models of Bianchi type I, one should consider
more general ``silent'' configurations by relaxing the vanishing of the
vorticity and the magnetic part of the Weyl tensor but maintaining their
``silence'' properties i.e. the vanishing of the curls of $E_{ab},H_{ab}$
and the pressure $p$.
\end{abstract}

\vskip 2cm

\clearpage

\section{Introduction}

\setcounter{equation}{0}

The discovery of the Szekeres family of solutions of the Einstein's Field
Equations (EFE), initiated a continuous research effort to understand the
richness of the structure of this interesting class of inhomogeneous models 
\cite{Szekeres1}-\cite{Goode-Wainwright}. The main reason for this interest
was that, due to the inhomogeneity condition, these solutions can in
principle describe either expanding or collapsing matter distributions
leading for example to a deeper insight into the formation mechanism of the
local structure inhomogeneities due to the amplification of small density
fluctuations in the early universe \cite{Bertschinger-Jain}.

Nevertheless, it emerged that Szekeres models were ``hiding'' a sufficient
number of crucial, in some sense, properties in their intrinsic structure.
At a geometrical level, it was proved immediately after their discovery that
they admit no isometries i.e. there are no non-trivial Killing Vector Fields
(KVFs) admitted by the spacetime manifold. Therefore one could reasonably
expect that the absence of KVFs should make Szekeres models quite general
within the class of inhomogeneous models. However, the well known
kinematical properties of the vanishing of the acceleration and the
vorticity of the fluid velocity (or, in the case of the vacuum subfamily of
models, the timelike vector field normal to the spatial hypersurfaces $t=$%
const.), indicate that these models are more special than expected. In
addition, Barnes and Rowlingson have shown that, apart from they being
algebraically special (namely of Petrov type D), the magnetic part $H_{ab}$
of the Weyl tensor vanishes \cite{Barnes-Rowlingson}. Consequently Szekeres
models belong to a wider family of models satisfying $H_{ab}=0$ which can
then be of the Petrov type I, D or O. This family contains, as special
cases, many well-known and important solutions like the standard
Friedmann-Lema\^{i}tre (FL) cosmological model, the Spatially Homogeneous
(SH) Bianchi type I models, as well as other cosmological or astrophysical
solutions of the EFE with sound physical interest \cite{Krasinski}.

Furthermore this result opened the possibility that these models may
represent the exact solution analogues of the inhomogeneous first order
scalar perturbations of the FL model, since it has been shown that the
magnetic part of the Weyl tensor does not contain scalar contributions at
the linear regime \cite{Goode}. Although it has been proved that this is not
true at second (or higher) order perturbations, the undertaken analysis for
models satisfying the conditions:

\begin{equation}
p=0=q^{a},\hspace{0.2cm}\pi _{ab}=0,\hspace{0.2cm}\dot{u}_{a}=0,\hspace{0.2cm%
}\omega _{ab}=0,\hspace{0.2cm}H_{ab}\equiv \frac{1}{2}\epsilon _{ac}^{%
\hspace{0.3cm}ef}C_{efbd}u^{c}u^{d}=0.  \label{Silent-Condition1}
\end{equation}
has uncovered a striking (and, from a mathematical point of view, very often
desirable) feature of the complete set of the EFE: the decoupling of the
spatial divergence and curl equations (or equivalently the \emph{constraints}
i.e. the fully projected derivatives normal to the timelike vector field $%
u^{a}$) from the evolution equations of the kinematical and dynamical
variables \cite{Matarrese-Pantano-Saez1}-\cite{Kofman-Pogosyan}. As a
result, there is no physical mechanism for gravity wave propagation ($%
H_{ab}=0$) or the presence of sound waves ($p=0$) between any nearby
timelike curves, thus leading to the name \emph{silent models}. Furthermore,
due to the above mentioned decoupling, the associated equations for the
physical variables are reduced to a set of first-order ordinary differential
equations (odes) which has been studied using methods from the theory of
dynamical systems with a view to analyzing the asymptotic behaviour of
either expanding or collapsing collisionless matter configurations \cite
{Bruni-Matarrese-Pantano1},\cite{Bruni-Matarrese-Pantano2}.

However, one must complement the above discussion by examining whether a
suitable set of initial data \emph{evolves consistently}, which will ensure
that Spatially Inhomogeneous Irrotational Silent (SIIS) models retain their
generality within the class of inhomogeneous solutions of the EFEs. This is
equivalent to demanding that the spatial divergence and curl equations
(encoded in the set of the initial data), are consistent with the evolution
equations hence, they are preserved identically along the timelike
congruence $u^{a}$ \emph{without} imposing new geometrical, kinematical or
dynamical restrictions. Although it has been stated that the constraints are
generally consistent \cite{Lesame-Dunsby-Ellis},\cite{Lesame-Ellis-Dunsby}
however, it has been demonstrated that this is not the case \cite
{Bonilla-Mars-Senovilla-Sopuerta-Vera}. In particular, the consistency
analysis for the SIIS of the Petrov type I has shown that the constraints,
coming from the silent conditions (\ref{Silent-Condition1}), are not trivial
and after repeated propagation along $u^{a}$, we expect to obtain an
infinite chain of non-trivial algebraic relations which allows one to
conjecture that \emph{there are no SIIS models of the Petrov type I} \cite
{ElstPhD,Maartens-Waves,Elst-Uggla-Lesame-Ellis-Maartens,Sopuerta}

Clearly, if we assume that the conjecture is true, the Szekeres family of
solutions keeps a privileged position within the SIIS models due to its
geometrical and dynamical characteristics and can be exploited in order to
clarify some open issues in spatially inhomogeneous models. For example, the
consistency of the Petrov type D models together with their non-symmetry
property (i.e. the non-existence of isometries), indicate that in the case
of the Petrov type I models the new constraints must be related with the
existence of some type of symmetry. This expectation can be justified by the
presence of the non-trivial algebraic (consistency) relations in the Petrov
type I models which necessary lead to a reduction of the allowed models,
similarly to the effect by the existence of a geometric symmetry. On the
other hand, the qualitative study of the SIIS models shows that, if the
Szekeres family of solutions is unique, we can restrict the dynamical
systems analysis only to the Petrov type D invariant set \cite
{Wainwright-Ellis-Book}. Among others, this provides a way to reveal the
geometric nature and to describe the dynamical role of the equilibrium
points of the associated state space and examine to what extent the
structure of the latter has similarities with the corresponding state space
of the SH models or the inhomogeneous models with one spatial degree of
freedom \cite{Wainwright-Ellis-Book}-\cite{vanElst:2001xm}.

The above discussion initiated a study towards either proving the conjecture
or, at least, presenting results that enforce its validity. For example, it
has been shown the uniqueness of the vacuum Szekeres sub-family within the
SIIS models and the non existence of Petrov type I dust models admitting a $%
G_{3}$ group of isometries, provided in both cases that the cosmological
constant $\Lambda $ vanishes \cite{Sopuerta,Mars},\cite
{VandenBergh-Wylleman2}. On the other hand, the presence of a cosmological
constant (i.e. the $\Lambda -$term generalization of the spatially
inhomogeneous and irrotational models studied in \cite{Barrow-Stein-Schabes}%
) cancels, in some way, the dynamical effects of the Petrov type I
conditions, therefore under certain circumstances the conjecture would not
hold and a broader family of models exists \cite{VandenBergh-Wylleman1}.

The purpose of this article is to revisit the consistency problem in vacuum
and pressureless SIIS models and provide an alternative method to resolve
it, using the time propagation of the constraints only up to third order and
avoiding the appearance of complicated algebraic relations between the
dynamical variables. The suggested approach is used to provide a conclusive
answer to the consistency conjecture in the affirmative, in a transparent
and fully covariant manner. Our method heavily relies on the use of the
theory of spacelike congruences in conjunction with the 1+3 covariant
formalism as applied in relativistic cosmology and astrophysics \cite
{Greenberg1}-\cite{Ellis-Elst1}. The paper is organized as follows: Section
2 is devoted to present the basic elements of the theory of timelike and
spacelike congruences and subsequently the 1+3 and 1+1+2 covariant
description of the SIIS models, in order to express the resultant
constraints in terms of the irreducible kinematical quantities of the
spacelike congruences. The main result of the paper is given in Section 3
where we prove the \emph{uniqueness} of the Petrov type D family within the
class of SIIS models. This is achieved by showing that models of the Petrov
type I \emph{always} admit a three dimensional Abelian group of isometries
with spacelike orbits, and thus reduce to the subclass of SH models of the
Bianchi type I \cite{Tsamp-Apostol4}. We should note that the above
uniqueness result is still valid when we allow the presence of a non-zero
pressure fluid configuration or a non-vanishing cosmological constant,
provided in both cases that the extrinsic curvature of the spacelike
hypersurfaces is not degenerated or, more generally, the shear eigenvalues
are not restricted to be proportional to the expansion of the timelike
congruence. Finally in Section 4 we draw our conclusions and discuss the
ways of how the approach presented in the present article can be extended to
check the broadness of more general silent models.

Throughout this paper, the following conventions have been used: the pair ($%
\mathcal{M},\mathbf{g}$) denotes the spacetime manifold endowed with a
Lorentzian metric of signature ($-,+,+,+$), spacetime indices are denoted by
lower case Latin letters $a,b,...=0,1,2,3$, spacelike eigenvalue indices are
denoted by lower case Greek letters $\alpha ,\beta ,...=1,2,3$ and we have
used geometrised units such that $8\pi G/c^2=1=c$.

\section{Covariant description of SIIS models}

\setcounter{equation}{0}

\subsection{Elements from the theory of timelike congruences}

The 1+3 covariant formalism \cite{Ellis-Elst1} starts by introducing a unit
timelike vector field $u^{a}$ ($u^{a}u_{a}=-1$) which is identified with the
average velocity of matter in the Universe, thus representing the congruence
of worldlines of the so-called fundamental (preferred) observers. The
existence of the timelike vector field $u^{a}$ generates, at any spacetime
event, a unique splitting of all the geometrical, kinematical and dynamical
(tensorial) quantities into temporal and spatial parts using the projection
tensor:\ 
\begin{equation}
h_{ab}\equiv g_{ab}+u_{a}u_{b},\hspace{0.5cm}h_{a}^{\hspace{0.2cm}%
c}h_{cb}=h_{ab},\hspace{0.5cm}h_{a}^{\hspace{0.2cm}b}h_{b}^{\hspace{0.2cm}%
a}=3,\hspace{0.5cm}h_{ab}u^{b}=0.  \label{projectiontensor}
\end{equation}
Essentially, $h_{ab}$ represents the intrinsic metric of the 3-dimensional
rest spaces of the observers $u^{a}$, even if these 3-spaces do not
constitute, in general, an integrable submanifold $\mathcal{S}$ of $\mathcal{%
M}$. The external geometric structure of the submanifold $\mathcal{S}$ can
be efficiently described in terms of the irreducible kinematical parts,
coming from the 1+3 decomposition of the first covariant derivatives of $%
u^{a}$ according to: 
\begin{equation}
u_{a;b}=\sigma _{ab}+\frac{\theta }{3}h_{ab}+\omega _{ab}-\dot{u}_{a}u_{b}
\label{timelikedecomposition1}
\end{equation}
where $\theta =u_{a;b}h^{ab}$ is the overall volume expansion (or
contraction) rate, $\sigma _{ab}=h_{a}^{\hspace{0.15cm}k}h_{b}^{\hspace{%
0.15cm}l}\left[ u_{\left( k;l\right) }-\frac{\theta }{3}h_{kl}\right] $ is
the shear tensor describing the rate of distortion of $\mathcal{S}$ in
different directions (i.e. the change of its shape), $\omega _{ab}=h_{a}^{%
\hspace{0.15cm}k}h_{b}^{\hspace{0.15cm}l}u_{\left[ k;l\right] }$ is the
vorticity tensor and $\dot{u}_{a}=u_{a;b}u^{b}$ is the four-acceleration
vector field. Moreover, since $u^{a}$ corresponds to the average velocity of
the matter fluid, the above kinematical quantities have also a direct
physical interpretation. Their dynamical behaviour can be studied using the
Ricci identity: 
\begin{equation}
2u_{a;[bc]}=R_{dabc}u^{d}  \label{RicciIdentity1}
\end{equation}
and taking into account the EFE.

For the case under consideration the fluid pressure vanishes and the EFE are
given by: 
\begin{equation}
R_{ab}=\frac{\rho }{2}\left( u_{a}u_{b}+h_{ab}\right) +\Lambda g_{ab}
\label{EFE1}
\end{equation}
where $\rho $ is the energy density of the matter fluid and $\Lambda$ is the
cosmological constant.

The complete 1+3 decomposition of (\ref{RicciIdentity1}) into temporal and
spatial parts, leads to \emph{evolution} and \emph{constraint} equations of
the kinematical quantities. Because the timelike congruence $u^{a}$ is
irrotational ($\omega _{ab}=0$) and geodesic ($\dot{u}_{a}=0$) and using the
``silent'' conditions (\ref{Silent-Condition1}), the Ricci identity reduces
to two evolution equations for $\theta $ and $\sigma _{ab}$ and two
constraint (divergence and curl) equations for the shear tensor \cite
{Ellis-Elst1}: 
\begin{equation}
\dot{\theta}=-\frac{1}{3}\theta ^{2}-\sigma _{ab}\sigma ^{ab}-\frac{1}{2}%
\rho +\Lambda  \label{evolutionexpansion}
\end{equation}
\begin{equation}
\dot{\sigma}_{ab}=-\frac{2}{3}\theta \sigma _{ab}-\sigma _{a}^{\hspace{0.2cm}%
c}\sigma _{cb}-E_{ab}+\frac{1}{3}\left( \sigma _{cd}\sigma ^{cd}\right)
h_{ab}  \label{evolutionshear}
\end{equation}
\begin{equation}
h_{ac}\sigma _{\hspace{0.3cm};k}^{kc}=\frac{2}{3}h_{a}^{\hspace{0.2cm}%
k}\theta _{;k}  \label{divergenceshear}
\end{equation}
\begin{equation}
h^{l(a}\epsilon ^{b)mns}\sigma _{nl;m}u_{s}=0  \label{shearconstraint}
\end{equation}
where 
\begin{equation}
E_{ab}\equiv C_{acbd}u^{c}u^{d}  \label{electricdefinition1}
\end{equation}
is the electric part of the Weyl conformal tensor and we have used the
notation: 
\begin{equation}
\dot{K}_{a...}\equiv K_{a...;k}u^{k}  \label{timederivdefinition}
\end{equation}
for the time derivative of any scalar or tensorial quantity.

We should note that the shear-curl constraint (\ref{shearconstraint}) is the
result of imposing the silent conditions (\ref{Silent-Condition1}). In order
to obtain a closed set of equations one must also use the Bianchi identities 
\begin{equation}
R_{ab[cd;e]}=0.  \label{BianchiIdentities1}
\end{equation}
The associated 1+3 decomposition of (\ref{BianchiIdentities1}) gives the
energy and momentum conservation equations plus two evolution and two
constraints equations for the electric and magnetic part of the Weyl tensor
which for the case of a SIIS model take the form: 
\begin{equation}
\dot{E}_{ab}=-\frac{1}{2}\rho \sigma _{ab}-\theta E_{ab}+3E_{(a}^{\hspace{%
0.2cm}c}\sigma _{b)c}-\left( E_{cd}\sigma ^{cd}\right) h_{ab}
\label{evolutionelectric}
\end{equation}
\begin{equation}
\dot{\rho}=-\rho \theta  \label{energyconservation}
\end{equation}
\begin{equation}
h_{ac}E_{\hspace{0.3cm};k}^{kc}=\frac{1}{3}h_{a}^{\hspace{0.2cm}k}\rho _{;k}
\label{electricdivergence}
\end{equation}
\begin{equation}
h^{l(a}\epsilon ^{b)mns}E_{nl;m}u_{s}=0  \label{electricconstraint}
\end{equation}
\begin{equation}
\epsilon ^{abcd}E_{bk}\sigma _{c}^{\hspace{0.2cm}k}u_{d}=0.
\label{magneticconstraint}
\end{equation}
Similarly, the electric-curl constraint (\ref{electricconstraint}) follows
from the silent conditions (\ref{Silent-Condition1}). The above closed
system of equations completely describes the dynamics and can be used in
order to determine the broadness of the family of SIIS models. This depends
on the consistency of the constraints (\ref{shearconstraint}) and (\ref
{electricconstraint}) with the set of evolution equations (\ref
{evolutionexpansion})-(\ref{evolutionshear}) and (\ref{evolutionelectric})-(%
\ref{energyconservation}). Although equations (\ref{shearconstraint}) and (%
\ref{electricconstraint}) appear similar they have a completely different
origin and one should expect that their consistency demands will be
independent. Eventually this is indeed the case since it has been shown that
the propagation of (\ref{shearconstraint}) does not generate further
conditions and evolves consistently along the worldlines of the fundamental
observers \cite{Maartens-Waves}. On the other hand propagation of (\ref
{electricconstraint}) leads to a set of non-trivially satisfied equations,
indicating that equation (\ref{electricconstraint}) represents a \emph{new}
constraint. It has been pointed out that the different character of (\ref
{electricconstraint}) is a reminiscent of the implications of the fact that
the vanishing of the evolution equation of a dynamical variable leads to a
new integrability condition \cite{Elst-Uggla-Lesame-Ellis-Maartens}. As we
shall demonstrate in the next section, equation (\ref{electricconstraint})
is not, in general, consistent with the evolution equations (\ref
{evolutionshear}) and (\ref{evolutionelectric})-(\ref{energyconservation})
and represents the compact form of a set of ``hidden'' geometric
restrictions.

A direct consequence of equation (\ref{magneticconstraint}) is that the
shear $\sigma _{ab}$ and the electric part $E_{ab}$ tensors commute which
implies that they have a common eigenframe. As a result, if $\left\{
x^{a},y^{a},z^{a}\right\} $ is a set of three mutually orthogonal and unit
spacelike vector fields that constitute the spatial eigenframe of $\sigma
_{ab}$ and $E_{ab}$, we may write:

\begin{equation}
E_{ab}=E_{1}x_{a}x_{b}+E_{2}y_{a}y_{b}+E_{3}z_{a}z_{b}
\label{electricparteigen}
\end{equation}
\begin{equation}
\sigma _{ab}=\sigma _{1}x_{a}x_{b}+\sigma _{2}y_{a}y_{b}+\sigma
_{3}z_{a}z_{b}  \label{sheareigen}
\end{equation}
where $E_{\alpha },\sigma _{\alpha }$ are the associated eigenvalues
satisfying the trace-free conditions: 
\begin{equation}
\sum_{\alpha }E_{\alpha }=\sum_{\alpha }\sigma _{\alpha }=0.
\label{trace-freecondition}
\end{equation}
Furthermore it has been shown that each of $\left\{
x^{a},y^{a},z^{a}\right\} $ is hypersurface orthogonal \cite
{Barnes-Rowlingson} i.e. 
\begin{equation}
x_{[a}x_{b;c]}=y_{[a}y_{b;c]}=z_{[a}z_{b;c]}=0.
\label{hypersurfaceorthogonal}
\end{equation}
which implies that $\dot{x}_{a}=\dot{y}_{a}=\dot{z}_{a}=0$ and local
coordinates can be found such that the metric can be written: 
\begin{equation}
ds^{2}=g_{ab}dx^{a}dx^{b}=-dt^{2}+e^{2A}dx^{2}+e^{2B}dy^{2}+e^{2\Gamma
}dz^{2}  \label{SIISmetric}
\end{equation}
where $A(t,x,y,z),B(t,x,y,z),\Gamma (t,x,y,z)$ are \emph{smooth} functions
of their arguments.

\subsection{Elements from the theory of spacelike congruences}

In many situations with clear geometrical or physical importance, apart from
the existence of a preferred timelike congruence, there also may exist a
preferred spacelike direction $x^a$ ($x^ax_a=1$) representing an intrinsic
geometrical or physical feature of the corresponding model (for example, due
to the existence of a spacelike KVF or the presence of a magnetic field).
Consequently, a further 1+2 splitting of the 3-dimensional space naturally
arises leading to the concept of the 1+1+2 decomposition of the spacetime
manifold.

In complete analogy with the 1+3 decomposition, the starting point is to
introduce the projection tensor: 
\begin{equation}
p_{ab}(\mathbf{x})\equiv g_{ab}+u_{a}u_{b}-x_{a}x_{b}=h_{ab}-x_{a}x_{b}
\label{xprojectiontensor}
\end{equation}
\begin{equation}
p_{a}^{\hspace{0.2cm}c}p_{cb}=p_{ab},\hspace{0.5cm}p_{a}^{\hspace{0.2cm}%
b}p_{b}^{\hspace{0.2cm}a}=2,\hspace{0.5cm}p_{ab}u^{b}=0=p_{ab}x^{b}
\label{xprojectionproperties}
\end{equation}
which is identified with the associated metric of the 2-dimensional space $%
\mathcal{X}$ (the so-called \emph{screen space}) normal to each vector field
of the pair $\left\{ u^{a},x^{a}\right\} $ at any spacetime event. Then the
geometric structure of $\mathcal{X}$ is studied by decomposing, into
irreducible kinematical parts, the first covariant derivatives of the
spacelike vector field $x^{a}$ according to \cite{Greenberg1}-\cite
{Saridakis-Tsamparlis}: 
\begin{equation}
x_{a;b}=\mathcal{T}_{ab}(\mathbf{x})+\frac{\mathcal{E}_{x}}{2}p_{ab}+%
\mathcal{R}_{ab}(\mathbf{x})+\stackrel{\ast }{x}_{a}x_{b}-\dot{x}%
_{a}u_{b}+p_{b}^{\hspace{0.2cm}c}\dot{x}_{c}u_{a}+\left[ 2\omega
_{cb}x^{c}-N_{b}(\mathbf{x})\right] u_{a}  \label{xderivativedecomposition1}
\end{equation}
where: 
\begin{equation}
\mathcal{E}_{x}=x_{a;b}p^{ab}=x_{\hspace{0.2cm};a}^{a}+\dot{x}^{a}u_{a}
\label{xexpansion}
\end{equation}
\begin{equation}
\mathcal{T}_{ab}(\mathbf{x})=p_{a}^{\hspace{0.15cm}k}p_{b}^{\hspace{0.15cm}l}%
\left[ x_{\left( k;l\right) }-\frac{1}{2}\mathcal{E}_{x}p_{kl}\right] ,%
\hspace{0.5cm}\mathcal{T}_{ab}p^{ab}=0  \label{xshear}
\end{equation}
\begin{equation}
\mathcal{R}_{ab}(\mathbf{x})=p_{a}^{\hspace{0.15cm}k}p_{b}^{\hspace{0.15cm}%
l}x_{[k;l]}  \label{xrotation}
\end{equation}
\begin{equation}
N_{a}=p_{a}^{\hspace{0.15cm}k}\left( \dot{x}_{k}-\stackrel{\ast }{u}%
_{k}\right)   \label{xgreenberg}
\end{equation}
are the rate of the surface expansion, the rate of shear tensor, the
rotation tensor and the Greenberg vector field of the spacelike congruence $%
x^{a}$ respectively and we have used the notation: 
\begin{equation}
\stackrel{\ast }{K}_{a...}\equiv K_{a...;k}x^{k}
\label{xderivativedefinition}
\end{equation}
for the directional derivative along the vector field $x^{a}$ of any scalar
or tensorial quantity.

The geometrical meaning of each of the defined kinematical quantities of the
spacelike congruence is similar to that of the corresponding quantities of $%
u^{a}$, carrying information on the (overall or in different directions)
distortion of $\mathcal{X}$ as measured by the fundamental observers $u^{a}$%
. The new ingredient is the appearance of the Greenberg vector $N_{a}$ which
is of crucial importance in the theory of spacelike congruences and
represents the ``coupling'' mechanism of the dynamical behaviour of the
model in directions normal and parallel to the screen space $\mathcal{X}$.
For example, the equation $N^{a}=0$ implies that the pair of vector fields $%
\left\{ u^{a},x^{a}\right\} $ generates a 2-dimensional integrable
submanifold of $\mathcal{M}$ and the spacelike congruence $x^{a}$ is
``comoving'' (``frozen-in'') along the worldlines of the fundamental
observers $u^{a}$. In addition this fact ensures that $\mathcal{T}_{ab}(%
\mathbf{x})$ and $\mathcal{R}_{ab}(\mathbf{x})$ lie in the screen space and
the unit vector fields $\left\{ x^{a},u^{a}\right\} $ are orthogonal at any
spacetime instant.

The incorporation of the EFE is achieved by applying (\ref{RicciIdentity1})
and (\ref{BianchiIdentities1}) to the unit vector field $x^{a}$. This leads
to propagation (along $x^{a}$) and ``constraint'' (lying on the screen
space) equations which describe the influence of the gravitational field in
the spatial variation of the kinematical and dynamical variables of the
timelike and spacelike congruences. In the next section we shall derive the
complete set of the dynamical equations associated with the orthonormal
tetrad $\{u^{a},x^{a},y^{a},z^{a}\}$ in order to analyze and express the
evolution and constraint equations (\ref{evolutionshear})-(\ref
{shearconstraint}) and (\ref{evolutionelectric})-(\ref{electricconstraint})
in terms of the kinematical quantities of the spacelike congruences with a
view to reveal the special structure of the new constraint (\ref
{electricconstraint}).

%%%%%%%%%%%%%%%%%%%%%%%%%%%%%%%%%%%%%%%%%%%%%%%%%%%%%

\subsection{SIIS models and spacelike congruences}

From equations (\ref{hypersurfaceorthogonal}) and after appropriate
projections, it follows that each of the eigenvectors $\left\{
x^{a},y^{a},z^{a}\right\} $ has the following properties:\ 
\begin{equation}
\dot{x}_{a}=N_{a}(\mathbf{x})=0=\mathcal{R}_{ab}(\mathbf{x})
\label{xproperties}
\end{equation}
\begin{equation}
\dot{y}_{a}=N_{a}(\mathbf{y})=0=\mathcal{R}_{ab}(\mathbf{y})
\label{yproperties}
\end{equation}
\begin{equation}
\dot{z}_{a}=N_{a}(\mathbf{z})=0=\mathcal{R}_{ab}(\mathbf{z})
\label{zproperties}
\end{equation}
which implies that each generator of the spacelike congruence is comoving
with the fundamental observers and the corresponding screen spaces are \emph{%
integrable} submanifolds of $\mathcal{M}$. Projecting equation (\ref
{shearconstraint}) or (\ref{electricconstraint}) with $x^{a}x^{b}$, $%
y^{a}y^{b}$ and $z^{a}z^{b}$ we obtain:\ 
\begin{equation}
\mathcal{T}_{ab}(\mathbf{x})=\alpha \left( y_{a}y_{b}-z_{a}z_{b}\right)
\label{xshearrelation}
\end{equation}
\begin{equation}
\mathcal{T}_{ab}(\mathbf{y})=\beta \left( z_{a}z_{b}-x_{a}x_{b}\right)
\label{yshearrelation}
\end{equation}
\begin{equation}
\mathcal{T}_{ab}(\mathbf{z})=\gamma \left( x_{a}x_{b}-y_{a}y_{b}\right)
\label{zshearrelation}
\end{equation}
where $\alpha ,\beta ,\gamma $ are the eigenvalues of the associated shear
tensors of the spacelike vector fields $x^{a},y^{a},z^{a}$ respectively.

Furthermore projecting (\ref{shearconstraint}) and (\ref{electricconstraint}%
) along $y^{a}z^{b}$, $z^{a}x^{b}$ and $x^{a}y^{b}$, the shear and electric
part constraints can be expressed as spatial variations of the corresponding
eigenvalues along the individual spacelike curve: 
\begin{equation}
\left( \sigma _{3}-\sigma _{2}\right) ^{\ast }=-\frac{\mathcal{E}_{x}}{2}%
\left( \sigma _{3}-\sigma _{2}\right) -3\alpha \sigma _{1}
\label{xshearconstraint}
\end{equation}
\begin{equation}
\left( \sigma _{1}-\sigma _{3}\right) ^{\prime }=-\frac{\mathcal{E}_{y}}{2}%
\left( \sigma _{1}-\sigma _{3}\right) -3\beta \sigma _{2}
\label{yshearconstraint}
\end{equation}
\begin{equation}
\left( \sigma _{2}-\sigma _{1}\right) ^{\symbol{126}}=-\frac{\mathcal{E}_{z}%
}{2}\left( \sigma _{2}-\sigma _{1}\right) -3\gamma \sigma _{3}
\label{zshearconstraint}
\end{equation}
\begin{equation}
\left( E_{3}-E_{2}\right) ^{\ast }=-\frac{\mathcal{E}_{x}}{2}\left(
E_{3}-E_{2}\right) -3\alpha E_{1}  \label{xelectricconstraint}
\end{equation}
\begin{equation}
\left( E_{1}-E_{3}\right) ^{\prime }=-\frac{\mathcal{E}_{y}}{2}\left(
E_{1}-E_{3}\right) -3\beta E_{2}  \label{yelectricconstraint}
\end{equation}
\begin{equation}
\left( E_{2}-E_{1}\right) ^{\symbol{126}}=-\frac{\mathcal{E}_{z}}{2}\left(
E_{2}-E_{1}\right) -3\gamma E_{3}  \label{zelectricconstraint}
\end{equation}
where we have introduced the notations: 
\begin{equation}
K_{a...}^{\prime }\equiv K_{a...;k}y^{k},\hspace{0.5cm}\left(
K_{a...}\right) ^{\symbol{126}}\equiv K_{a...;k}z^{k}.
\label{yandzderivativedefinition}
\end{equation}
The advantage of decomposing the constraints along the specific spacelike
curve is immediately apparent from equations (\ref{xshearconstraint})-(\ref
{zelectricconstraint}): the associated spatial variations of the shear and
electric part eigenvalues have been partially decoupled and expressed in
terms of the corresponding kinematical variables of the spacelike
congruences. Obviously, this decoupling process is the result of the
conditions (\ref{xproperties})-(\ref{zproperties}).

The above set of equations must be augmented with the evolution equations (%
\ref{evolutionshear}), (\ref{evolutionelectric}), (\ref{energyconservation})
and the divergence equations (\ref{divergenceshear}), (\ref
{electricdivergence}):

\begin{equation}
\dot{\sigma}_{\alpha }=-\frac{2}{3}\theta \sigma _{\alpha }-\left( \sigma
_{\alpha }\right) ^{2}-E_{\alpha }+\frac{1}{3}\sum_{\alpha }\left( \sigma
_{\alpha }\right) ^{2}  \label{sheareigenvaluesevolution}
\end{equation}
\begin{equation}
\dot{E}_{\alpha }=-\frac{1}{2}\rho \sigma _{\alpha }-\theta E_{\alpha
}+3\sigma _{\alpha }E_{\alpha }-\sum_{\alpha }E_{\alpha }\sigma _{\alpha }
\label{electriceigenvaluesevolution}
\end{equation}
\begin{equation}
\dot{\rho}=-\rho \theta .  \label{energyconservation2}
\end{equation}
\begin{equation}
\stackrel{\ast }{\rho }=3\left[ \stackrel{\ast }{E_{1}}+\frac{3\mathcal{E}%
_{x}}{2}E_{1}-\alpha \left( E_{2}-E_{3}\right) \right] ,\hspace{0.5cm}%
\stackrel{\ast }{\theta }=\frac{3}{2}\left[ \stackrel{\ast }{\sigma _{1}}+%
\frac{3\mathcal{E}_{x}}{2}\sigma _{1}-\alpha \left( \sigma _{2}-\sigma
_{3}\right) \right]  \label{xdivergence}
\end{equation}
\begin{equation}
\rho ^{\prime }=3\left[ E_{2}^{\prime }+\frac{3\mathcal{E}_{y}}{2}%
E_{2}-\beta \left( E_{3}-E_{1}\right) \right] ,\hspace{0.5cm}\theta ^{\prime
}=\frac{3}{2}\left[ \sigma _{2}^{\prime }+\frac{3\mathcal{E}_{y}}{2}\sigma
_{2}-\beta \left( \sigma _{3}-\sigma _{1}\right) \right]  \label{ydivergence}
\end{equation}
\begin{equation}
\tilde{\rho}=3\left[ \tilde{E}_{3}+\frac{3\mathcal{E}_{z}}{2}E_{3}-\gamma
\left( E_{1}-E_{2}\right) \right] ,\hspace{0.5cm}\tilde{\theta}=\frac{3}{2}%
\left[ \tilde{\sigma}_{3}+\frac{3\mathcal{E}_{z}}{2}\sigma _{3}-\gamma
\left( \sigma _{1}-\sigma _{2}\right) \right] .  \label{zdivergence}
\end{equation}
Using a similar splitting of the Ricci identities for each spacelike vector
field of the orthonormal triad $\left\{ x^{a},y^{a},z^{a}\right\} $, we
obtain evolution (along $u^{a}$), propagation (along the individual
spacelike vector field) and ``constraint'' (lying in the associated screen
space) equations of the kinematical quantities of the spacelike congruences
which we give them below.

The \emph{evolution} and the ``\emph{constraint}'' equations arise from the
temporal and the spatial projections of the trace part of the Ricci identity
respectively:

\begin{equation}
\left( \mathcal{E}_{x}\right) ^{\cdot }=\frac{1}{2}\mathcal{E}_{x}\sigma
_{1}-\frac{1}{3}\mathcal{E}_{x}\theta -\alpha \left( \sigma _{2}-\sigma
_{3}\right) ,\hspace{0.5cm}\dot{\alpha}=\frac{1}{2}\alpha \sigma _{1}-\frac{1%
}{3}\alpha \theta -\frac{1}{4}\mathcal{E}_{x}\left( \sigma _{2}-\sigma
_{3}\right)  \label{evolutionxcongruence}
\end{equation}
\begin{equation}
\left( \mathcal{E}_{y}\right) ^{\cdot }=\frac{1}{2}\mathcal{E}_{y}\sigma
_{2}-\frac{1}{3}\mathcal{E}_{y}\theta -\beta \left( \sigma _{3}-\sigma
_{1}\right) ,\hspace{0.5cm}\dot{\beta}=\frac{1}{2}\beta \sigma _{2}-\frac{1}{%
3}\beta \theta -\frac{1}{4}\mathcal{E}_{y}\left( \sigma _{3}-\sigma
_{1}\right)  \label{evolutionycongruence}
\end{equation}
\begin{equation}
\left( \mathcal{E}_{z}\right) ^{\cdot }=\frac{1}{2}\mathcal{E}_{z}\sigma
_{3}-\frac{1}{3}\mathcal{E}_{z}\theta -\gamma \left( \sigma _{1}-\sigma
_{2}\right) ,\hspace{0.5cm}\dot{\gamma}=\frac{1}{2}\gamma \sigma _{3}-\frac{1%
}{3}\gamma \theta -\frac{1}{4}\mathcal{E}_{z}\left( \sigma _{1}-\sigma
_{2}\right)  \label{evolutionzcongruence}
\end{equation}
\begin{equation}
\left( \frac{\mathcal{E}_{x}}{2}-\alpha \right) ^{\prime }=2\alpha \left( 
\frac{\mathcal{E}_{y}}{2}+\beta \right) ,\hspace{0.5cm}\left( \frac{\mathcal{%
E}_{x}}{2}+\alpha \right) ^{\symbol{126}}=-2\alpha \left( \frac{\mathcal{E}%
_{z}}{2}-\gamma \right)  \label{spatialXRicciIdentity}
\end{equation}
\begin{equation}
\left( \frac{\mathcal{E}_{y}}{2}+\beta \right) ^{\ast }=-2\beta \left( \frac{%
\mathcal{E}_{x}}{2}-\alpha \right) ,\hspace{0.5cm}\left( \frac{\mathcal{E}%
_{y}}{2}-\beta \right) ^{\symbol{126}}=2\beta \left( \frac{\mathcal{E}_{z}}{2%
}+\gamma \right)  \label{spatialYRicciIdentity}
\end{equation}
\begin{equation}
\left( \frac{\mathcal{E}_{z}}{2}-\gamma \right) ^{\ast }=2\gamma \left( 
\frac{\mathcal{E}_{x}}{2}+\alpha \right) ,\hspace{0.5cm}\left( \frac{%
\mathcal{E}_{z}}{2}+\gamma \right) ^{\prime }=-2\gamma \left( \frac{\mathcal{%
E}_{y}}{2}-\beta \right) .  \label{spatialZRicciIdentity}
\end{equation}
It is interesting to note that the evolution equations (\ref
{evolutionxcongruence})-(\ref{evolutionzcongruence}) imply for example $%
\alpha =0\Leftrightarrow \mathcal{E}_{x}=0$ for a Petrov type I model. In
addition, by taking appropriate linear combinations of the ``constraint''
equations (\ref{spatialXRicciIdentity})-(\ref{spatialZRicciIdentity}) this
set is essentially equivalent with the Jacobi identities or the twist-free
property $\mathcal{R}=0$ of the spacelike congruences.

The \emph{propagation} equations of the expansion and the shear of the
spacelike curves follow from the trace and trace free part of the Ricci
identity \cite{Tsamparlis-Mason} which in the case of the SIIS models take
the form:

\begin{eqnarray}
(\mathcal{E}_{x})^{\ast } &=&-\frac{\mathcal{E}_{x}^{2}}{2}-2\alpha
^{2}-\left( E_{1}+\frac{2\rho }{3}\right) +\left( \sigma _{1}+\frac{\theta }{%
3}\right) \left( \frac{2\theta }{3}-\sigma _{1}\right)  \nonumber \\
&&-\left( \frac{\mathcal{E}_{z}}{2}+\gamma \right) ^{\symbol{126}}-\left( 
\frac{\mathcal{E}_{z}}{2}+\gamma \right) \mathcal{E}_{z}-\left( \frac{%
\mathcal{E}_{y}}{2}-\beta \right) ^{\prime }-\left( \frac{\mathcal{E}_{y}}{2}%
-\beta \right) \mathcal{E}_{y}  \label{xpropagationequationexpansion}
\end{eqnarray}
\begin{eqnarray}
(\mathcal{E}_{y})^{\prime } &=&-\frac{\mathcal{E}_{y}^{2}}{2}-2\beta
^{2}-\left( E_{2}+\frac{2\rho }{3}\right) +\left( \sigma _{2}+\frac{\theta }{%
3}\right) \left( \frac{2\theta }{3}-\sigma _{2}\right)  \nonumber \\
&&-\left( \frac{\mathcal{E}_{x}}{2}+\alpha \right) ^{\ast }-\left( \frac{%
\mathcal{E}_{x}}{2}+\alpha \right) \mathcal{E}_{x}-\left( \frac{\mathcal{E}%
_{z}}{2}-\gamma \right) ^{\symbol{126}}-\left( \frac{\mathcal{E}_{z}}{2}%
-\gamma \right) \mathcal{E}_{z}  \label{ypropagationequationexpansion}
\end{eqnarray}
\begin{eqnarray}
(\mathcal{E}_{z})^{\symbol{126}} &=&-\frac{\mathcal{E}_{z}^{2}}{2}-2\gamma
^{2}-\left( E_{3}+\frac{2\rho }{3}\right) +\left( \sigma _{3}+\frac{\theta }{%
3}\right) \left( \frac{2\theta }{3}-\sigma _{3}\right)  \nonumber \\
&&-\left( \frac{\mathcal{E}_{y}}{2}+\beta \right) ^{\prime }-\left( \frac{%
\mathcal{E}_{y}}{2}+\beta \right) \mathcal{E}_{y}-\left( \frac{\mathcal{E}%
_{x}}{2}-\alpha \right) ^{\ast }-\left( \frac{\mathcal{E}_{x}}{2}-\alpha
\right) \mathcal{E}_{x}  \label{zpropagationequationexpansion}
\end{eqnarray}
\begin{eqnarray}
\left( \alpha \right) ^{\ast } &=&-\alpha \mathcal{E}_{x}-\frac{1}{2}\left( 
\frac{\mathcal{E}_{y}}{2}-\beta \right) ^{\prime }+\left( \frac{\mathcal{E}%
_{y}}{2}-\beta \right) \beta +\frac{1}{2}\left( \frac{\mathcal{E}_{z}}{2}%
+\gamma \right) ^{\symbol{126}}+\left( \frac{\mathcal{E}_{z}}{2}+\gamma
\right) \gamma  \nonumber \\
&&-\frac{E_{1}+2E_{2}}{2}+\frac{1}{2}\left( \sigma _{1}+2\sigma _{2}\right)
\left( \sigma _{1}+\frac{\theta }{3}\right)
\label{xpropagationequationshear}
\end{eqnarray}
\begin{eqnarray}
\left( \beta \right) ^{\prime } &=&-\beta \mathcal{E}_{y}-\frac{1}{2}\left( 
\frac{\mathcal{E}_{z}}{2}-\gamma \right) ^{\symbol{126}}+\left( \frac{%
\mathcal{E}_{z}}{2}-\gamma \right) \gamma +\frac{1}{2}\left( \frac{\mathcal{E%
}_{x}}{2}+\alpha \right) ^{\ast }+\left( \frac{\mathcal{E}_{x}}{2}+\alpha
\right) \alpha  \nonumber \\
&&-\frac{E_{2}+2E_{3}}{2}+\frac{1}{2}\left( \sigma _{2}+2\sigma _{3}\right)
\left( \sigma _{2}+\frac{\theta }{3}\right)
\label{ypropagationequationshear}
\end{eqnarray}
\begin{eqnarray}
\left( \gamma \right) ^{\symbol{126}} &=&-\gamma \mathcal{E}_{z}-\frac{1}{2}%
\left( \frac{\mathcal{E}_{x}}{2}-\alpha \right) ^{\ast }+\left( \frac{%
\mathcal{E}_{x}}{2}-\alpha \right) \alpha +\frac{1}{2}\left( \frac{\mathcal{E%
}_{y}}{2}+\beta \right) ^{\prime }+\left( \frac{\mathcal{E}_{y}}{2}+\beta
\right) \beta  \nonumber \\
&&-\frac{E_{3}+2E_{1}}{2}+\frac{1}{2}\left( \sigma _{3}+2\sigma _{1}\right)
\left( \sigma _{3}+\frac{\theta }{3}\right)
\label{zpropagationequationshear}
\end{eqnarray}
The set of equations (\ref{xshearconstraint})-(\ref{zelectricconstraint})
and (\ref{sheareigenvaluesevolution})-(\ref{zpropagationequationshear})
completely characterizes the dynamics of SIIS models and it will be used in
order to prove the conjecture by showing that vacuum and non-vacuum silent
models of Petrov type I \emph{always admit a three dimensional group of
isometries} thus reducing to the subclass of SH models of the Bianchi type I.

Finally it should be remarked that there exists a direct correspondence
between the description of SIIS models with the theory of spacelike
congruences and the Orthonormal Frame (ONF) approach \cite
{Ellis-Elst1,Elst-Uggla}. For example, the spatial curvature quantities $%
n_{12},n_{31},n_{23},a_{1},a_{2},a_{3}$ and the kinematical quantities of
the spacelike congruences are related via:\ 
\begin{equation}
n_{23}=-\alpha ,\hspace{0.5cm}n_{31}=-\beta ,\hspace{0.5cm}n_{12}=-\gamma
\label{n-shearrelations}
\end{equation}
\begin{equation}
a_{1}=-\frac{\mathcal{E}_{x}}{2},\hspace{0.5cm}a_{2}=-\frac{\mathcal{E}_{y}}{%
2},\hspace{0.5cm}a_{3}=-\frac{\mathcal{E}_{z}}{2}
\label{a-expansionrelations}
\end{equation}
whereas the set (\ref{xshearconstraint})-(\ref{zelectricconstraint}) and (%
\ref{sheareigenvaluesevolution})-(\ref{zpropagationequationshear}) is the
covariant version of the tetrad equations presented in \cite
{Elst-Uggla-Lesame-Ellis-Maartens}. As we shall see in the next section,
this fact (already emphasized in \cite{Elst-Uggla-Lesame-Ellis-Maartens})
makes clear the covariant property of the incompatibility of the constraints
(\ref{xshearconstraint})-(\ref{zelectricconstraint}) with the evolution
equations. At the same time reveals in a transparent way, the \emph{%
geometric nature} of (\ref{xshearconstraint})-(\ref{zelectricconstraint})
since they induce, through the kinematical variables of the spacelike
congruences, a set of symmetry constraints leading to the subclass of SH
models of Bianchi type I.

%%%%%%%%%%%%%%%%%%%%%%%%%%%%%%%%%%%%%%%%%%%%%%%%%%%%%%%%%%%%%%%%%%%%%%%%%%%%%%%%%%%%%%%%%

\section{Uniqueness of the Szekeres solutions within the family of SIIS
models}

\setcounter{equation}{0}

As we have mentioned in the Introduction, the vanishing of the magnetic part
of the Weyl tensor gives rise to a set of constraints which in the 1+1+2
covariant formalism takes the form of equations (\ref{xshearconstraint})-(%
\ref{zelectricconstraint}). In order to be consistent with the evolution
equations (\ref{sheareigenvaluesevolution})-(\ref{energyconservation2}),
they must \emph{not} generate a new set of constraints after their repeated
propagation along the timelike congruence $u^{a}$. The outcome of this
procedure has led to the proof of the consistency of Petrov type D models
and that type I models develop a chain of \emph{new} highly non-linear and
non-trivial algebraic constraints that need to be solved \cite
{ElstPhD,Elst-Uggla-Lesame-Ellis-Maartens,Sopuerta}.

However, it will be enlightening if we choose an alternative direction to
deal with the consistency checking, in the spirit of the notion of geometric
symmetries. The concept of a geometric symmetry or collineation admitted by
the spacetime manifold, is closely related to that of a transformation that
leaves a specific geometric object invariant (up to a conformal factor)
along the integral lines of the generating vector field. Although there
exists (and can be defined) a sufficiently large number of geometric
symmetries, they have, in general, the drawback of imposing severe
restrictions on the geometry and the dynamics leading, in most cases, either
to physically unsound models or models with very special properties (see for
example \cite{Tsamp-Apostol4} for SH models of Bianchi type I). In order to
demonstrate this fact, let us consider the case of a spacelike Conformal
Vector Field (CVF) $X^{a}=Xx^{a}$ which is defined according to the relation:

\begin{equation}
\mathcal{L}_{\mathbf{X}}g_{ab}=2\psi g_{ab}  \label{ConformalCondition1}
\end{equation}
where $\psi $ is the scale amplification or dilation of the spacetime metric.

In terms of the kinematical variables of the spacelike congruence the
conformal equation (\ref{ConformalCondition1}) can be shown to be equivalent
to the conditions \cite{Saridakis-Tsamparlis}: 
\[
\mathcal{T}_{ab}(\mathbf{x})=0,\hspace{0.5cm}\left( \ln X\right) _{;a}=\frac{%
1}{2}\mathcal{E}_{x}x_{a}-\stackrel{\ast }{x}_{a}
\]
\begin{equation}
\dot{x}_{a}u^{a}=-\frac{1}{2}\mathcal{E}_{x},\hspace{0.5cm}N_{a}=-2\omega
_{ab}x^{b}  \label{ConformalCondition2}
\end{equation}
and the conformal factor is then given by $\psi =\frac{1}{2}X\mathcal{E}_{x}=%
\stackrel{\ast }{X}$.

Furthermore, the coupling of the geometry with the matter fields, through
the EFE, implies that there is a mutual influence between any geometrical or
dynamical constraint with subsequent restrictions in the structure of the
corresponding models. For example, in the case of an irrotational and
geodesic fluid flow the conditions (\ref{ConformalCondition2}) become:\ 
\[
\mathcal{T}_{ab}(\mathbf{x})=0,\hspace{0.5cm}\left( \ln X\right) _{;a}=-%
\stackrel{\ast }{x}_{a} 
\]
\begin{equation}
\mathcal{E}_{x}=0,\hspace{0.5cm}N_{a}=0,\hspace{0.5cm}\psi =0=\stackrel{\ast 
}{X}  \label{ConformalCondition3}
\end{equation}
i.e. the spacelike congruence is necessarily comoving with the observers $%
u^{a} $ and the CVF reduces to a KVF. Using the covariant description of the
Petrov type D models in terms of the kinematical quantities of the spacelike
congruences (see below), this in particular implies that the Szekeres
solutions \emph{cannot} admit spacelike CVF parallel to any of the
eigenvectors of the shear (or the Weyl electric part) tensor.

The consistency of the Petrov type D models along with their non-symmetry
property (non-existence of isometries), allows one to speculate about the
existence of some type of link between the consistency conditions and the
presence of a geometric symmetry in the case of the Petrov type I models.
This expectation can be justified by the fact that the new algebraic
constraints in Petrov type I models are coming from the time propagation of
the initial integrability conditions (\ref{xelectricconstraint})-(\ref
{zelectricconstraint}). In addition, they involve the kinematical and
dynamical variables and \emph{necessarily}\footnote{%
Even if we assume that at most one, in the whole set of algebraic relations,
is linearly independent this means that one of the kinematical or dynamical
variables of the model can be expressed as a function of the other, thus
leading to an associated restriction of their broadness.} lead to a
restriction of the underlying geometric structure of the corresponding
models. On the other hand the existence of any geometric symmetry is
controlled by a set of integrability conditions which, after repeated
covariant differentiation, also lead to a system of algebraic relations.
Consequently one may argue that the presence of the non-trivial constraints (%
\ref{xelectricconstraint})-(\ref{zelectricconstraint}) (together with the
rest of conditions imposed by (\ref{Silent-Condition1})) in SIIS models of
Petrov type I can be loosely interpreted as corresponding to some type of
symmetry restriction.\footnote{%
The interpretation of the induced constraints in terms of a set of symmetry
conditions indicates a possible connection between the full set of EFE and
the existence of a specific symmetry. For a subclass of SH vacuum and
non-titled perfect fluid models, this connection has been established
showing that a specific symmetry may play a role in the invariant
description of general cosmological models \cite
{Apostol-Tetrad-Appr-SH-Models}.} In the following we shall prove that this
is the case and the corresponding symmetry constraints are equivalent to the
Killing conditions (\ref{ConformalCondition3}).

The above arguments and the conclusions they lead to, can be summarized in
the following: \newline
\newline
\textbf{Theorem} \emph{There do not exist vacuum and non-vacuum SIIS models
of the Petrov type I with vanishing cosmological constant}\newline
\newline
Before we proceed with the proof of the Theorem, it will be useful to
covariantly characterize the SIIS models of the Petrov type D ($S_{D}$) in
terms of the kinematical variables of the spacelike congruences. Assuming
that $E_{1}=E_{2}\Leftrightarrow \sigma _{1}=\sigma _{2}$, equation (\ref
{zshearconstraint}) (or (\ref{zelectricconstraint})) implies that $\gamma
=0\Leftrightarrow \mathcal{T}_{ab}(\mathbf{z})=0\Leftrightarrow p_{a}^{%
\hspace{0.15cm}k}(\mathbf{z})p_{b}^{\hspace{0.15cm}l}(\mathbf{z})z_{k;l}=%
\frac{\mathcal{E}_{z}}{2}p_{ab}(\mathbf{z})$. Therefore SIIS models of the
Petrov type D are described by the conditions: 
\begin{equation}
E_{1}=E_{2}\Leftrightarrow \sigma _{1}=\sigma _{2},\hspace{0.5cm}\mathcal{T}%
_{ab}(\mathbf{z})=0,\hspace{0.5cm}  \label{SzekeresConditions}
\end{equation}
It is interesting to note that under the conditions (\ref{SzekeresConditions}%
) the complete set of equations (\ref{xshearconstraint})-(\ref
{zelectricconstraint}) and (\ref{sheareigenvaluesevolution})-(\ref
{evolutionzcongruence}) is consistent under repeated propagation along the
timelike congruence $u^{a}$. Furthermore, using the local coordinate form of
the SIIS metric (\ref{SIISmetric}) and choosing $x^{a}=e^{-A}\delta _{x}^{a}$%
, $y^{a}=e^{-B}\delta _{y}^{a}$, $z^{a}=e^{-\Gamma }\delta _{z}^{a}$ we can
verify that the conditions (\ref{SzekeresConditions}) follow from the
assumption $A=B+H$ where the function $H$ satisfies $\dot{H}=\tilde{H}=0$.%
\newline
\newline
\newline
\textbf{Proof}\newline
%%%%%%%%%%%%%%%%%%%%%%%%%%%%%%%%%%%%%%%%%%%%%%%%%%%%%%%%%%%%%%%%%%%%
Consider the case of Petrov type I models ($S_{I}$) which implies that $%
E_{\alpha }\neq E_{\beta }$ and $\sigma _{\alpha }\neq \sigma _{\beta }$ for
every $\alpha ,\beta =1,2,3$. Using the Ricci identity (\ref{RicciIdentity1}%
) it is straightforward to show the following commutation relations for
every scalar quantity $S$ (see also equation (A8) in \cite{Maartens-Waves}): 
\begin{equation}
\left( \stackrel{\ast }{S}\right) ^{\cdot }=\left( \dot{S}\right) ^{\ast }-%
\stackrel{\ast }{S}\left( \sigma _{1}+\frac{\theta }{3}\right)
\label{xcommutaionrelation}
\end{equation}
\begin{equation}
\left( S^{\prime }\right) ^{\cdot }=\left( \dot{S}\right) ^{\prime
}-S^{\prime }\left( \sigma _{2}+\frac{\theta }{3}\right)
\label{ycommutaionrelation}
\end{equation}
\begin{equation}
\left( \tilde{S}\right) ^{\cdot }=\left( \dot{S}\right) ^{\symbol{126}}-%
\tilde{S}\left( \sigma _{3}+\frac{\theta }{3}\right) .
\label{zcommutaionrelation}
\end{equation}
In order to analyzing the consistency of the constraints (\ref
{xelectricconstraint})-(\ref{zelectricconstraint}), we propagate them along $%
u^{a}$ and use the commutation relations (\ref{xcommutaionrelation})-(\ref
{zcommutaionrelation}) and the set of equations (\ref
{sheareigenvaluesevolution})-(\ref{electriceigenvaluesevolution}) and (\ref
{xdivergence})-(\ref{evolutionzcongruence}). As expected, the dynamical
conditions (\ref{xelectricconstraint})-(\ref{zelectricconstraint}) are not
identically satisfied and they lead to the following \emph{new} constraints: 
\begin{eqnarray}
6\left[ \stackrel{\ast }{E}_{2}\left( \sigma _{3}-\sigma _{2}\right) +%
\stackrel{\ast }{\sigma }_{2}\left( E_{3}-E_{2}\right) \right] &=&3\mathcal{E%
}_{x}\left[ E_{2}\left( 2\sigma _{2}-\sigma _{3}\right) -E_{3}\sigma _{2}%
\right]  \nonumber \\
&&  \nonumber \\
&&+2\alpha \left[ E_{2}\left( 2\sigma _{2}-11\sigma _{3}\right) -E_{3}\left(
11\sigma _{2}+16\sigma _{3}\right) \right]  \label{newXconstraint}
\end{eqnarray}
\begin{eqnarray}
6\left[ E_{3}^{\prime }\left( \sigma _{3}-\sigma _{1}\right) +\sigma
_{3}^{\prime }\left( E_{3}-E_{1}\right) \right] &=&3\mathcal{E}_{y}\left[
-E_{3}\left( 2\sigma _{3}-\sigma _{1}\right) +E_{1}\sigma _{3}\right] 
\nonumber \\
&&  \nonumber \\
&&+2\beta \left[ -E_{3}\left( 2\sigma _{3}-11\sigma _{1}\right) +E_{1}\left(
11\sigma _{3}+16\sigma _{1}\right) \right]  \label{newYconstraint}
\end{eqnarray}
\begin{eqnarray}
6\left[ \tilde{E}_{1}\left( \sigma _{2}-\sigma _{1}\right) +\tilde{\sigma}%
_{1}\left( E_{2}-E_{1}\right) \right] &=&3\mathcal{E}_{z}\left[ E_{1}\left(
2\sigma _{1}-\sigma _{2}\right) -E_{2}\sigma _{1}\right]  \nonumber \\
&&  \nonumber \\
&&+2\gamma \left[ E_{1}\left( 2\sigma _{1}-11\sigma _{2}\right) -E_{2}\left(
11\sigma _{1}+16\sigma _{2}\right) \right]  \label{newZconstraint}
\end{eqnarray}
Clearly, equations (\ref{newXconstraint})-(\ref{newZconstraint}) represent
the \emph{covariant} form of the constraints (67)-(69) given in \cite
{Elst-Uggla-Lesame-Ellis-Maartens}. Moreover, the assumption of a Petrov
type D model (for example $E_{1}=E_{2}\Leftrightarrow \sigma _{1}=\sigma
_{2} $) implies that $\gamma =0$ and the rest of equations are trivially
satisfied.

Solving equations (\ref{newXconstraint})-(\ref{newZconstraint}) with respect
to $\stackrel{\ast }{E}_{2}$, $E_{3}^{\prime }$, $\tilde{E}_{1}$, a second
temporal propagation leads to the following expressions for the spatial
derivatives $\stackrel{\ast }{\sigma }_{2}$, $\sigma _{3}^{\prime }$ and $%
\tilde{\sigma}_{1}$: 
\begin{equation}
\stackrel{\ast }{\sigma }_{2}=\frac{A_{1}}{A_{2}},\hspace{0.4cm}\sigma
_{3}^{\prime }=\frac{B_{1}}{B_{2}},\hspace{0.4cm}\tilde{\sigma}_{1}=\frac{%
C_{1}}{C_{2}}  \label{spatialderivesofshear}
\end{equation}
where: 
\begin{eqnarray*}
A_{1} &=&-2E_{2}^{2}\left[ 3\mathcal{E}_{x}\sigma _{2}+2\alpha \left( \sigma
_{2}-10\sigma _{3}\right) \right] \\
&&+E_{2}\{4E_{3}\left[ 3\mathcal{E}_{x}\sigma _{2}+2\alpha \left( \sigma
_{2}+8\sigma _{3}\right) \right] \\
&&+3\left( \sigma _{3}-\sigma _{2}\right) \left[ \mathcal{E}_{x}\left(
10\sigma _{2}^{2}+4\sigma _{2}\sigma _{3}-5\sigma _{3}^{2}\right) +2\alpha
\left( 14\sigma _{2}^{2}+20\sigma _{2}\sigma _{3}+11\sigma _{3}^{2}\right) %
\right] \} \\
&&-2E_{3}^{2}\left[ 3\mathcal{E}_{x}\sigma _{2}+2\alpha \left( 19\sigma
_{2}+8\sigma _{3}\right) \right] \\
&&-3E_{3}\left( \mathcal{E}_{x}+2\alpha \right) \left( \sigma _{2}-\sigma
_{3}\right) \left( 2\sigma _{2}^{2}-7\sigma _{2}\sigma _{3}-4\sigma
_{3}^{2}\right) \\
&&-\rho \left( \sigma _{3}-\sigma _{2}\right) \left[ 3\mathcal{E}_{x}\sigma
_{2}\left( \sigma _{2}-\sigma _{3}\right) +2\alpha \left( \sigma
_{2}^{2}-11\sigma _{2}\sigma _{3}-8\sigma _{3}^{2}\right) \right] \\
&& \\
A_{2} &=&6\{2E_{2}^{2}-E_{2}\left[ 4E_{3}+3\left( \sigma _{3}-\sigma
_{2}\right) \left( 2\sigma _{2}+\sigma _{3}\right) \right] \\
&&+2E_{3}^{2}+3E_{3}\left( \sigma _{3}-\sigma _{2}\right) \left( 2\sigma
_{3}+\sigma _{2}\right) -\rho \left( \sigma _{2}-\sigma _{3}\right) ^{2}]\}
\end{eqnarray*}

\bigskip

\begin{eqnarray*}
B_{1} &=&-2E_{3}^{2}\left[ 3\mathcal{E}_{y}\sigma _{3}+2\beta \left( \sigma
_{3}-10\sigma _{1}\right) \right] \\
&&+E_{3}\{4E_{1}\left[ 3\mathcal{E}_{y}\sigma _{3}+2\beta \left( \sigma
_{3}+8\sigma _{1}\right) \right] \\
&&+3\left( \sigma _{1}-\sigma _{3}\right) \left[ \mathcal{E}_{y}\left(
10\sigma _{3}^{2}+4\sigma _{1}\sigma _{3}-5\sigma _{1}^{2}\right) +2\beta
\left( 14\sigma _{3}^{2}+20\sigma _{1}\sigma _{3}+11\sigma _{1}^{2}\right) %
\right] \} \\
&&-2E_{1}^{2}\left[ 3\mathcal{E}_{y}\sigma _{3}+2\beta \left( 19\sigma
_{3}+8\sigma _{1}\right) \right] \\
&&-3E_{1}\left( \mathcal{E}_{y}+2\beta \right) \left( \sigma _{3}-\sigma
_{1}\right) \left( 2\sigma _{3}^{2}-7\sigma _{1}\sigma _{3}-4\sigma
_{1}^{2}\right) \\
&&-\rho \left( \sigma _{1}-\sigma _{3}\right) \left[ 3\mathcal{E}_{y}\sigma
_{3}\left( \sigma _{3}-\sigma _{1}\right) +2\beta \left( \sigma
_{3}^{2}-11\sigma _{1}\sigma _{3}-8\sigma _{1}^{2}\right) \right] \\
&& \\
B_{2} &=&6\{2E_{3}^{2}-E_{3}\left[ 4E_{1}+3\left( \sigma _{1}-\sigma
_{3}\right) \left( 2\sigma _{3}+\sigma _{1}\right) \right] \\
&&+2E_{1}^{2}+3E_{1}\left( \sigma _{1}-\sigma _{3}\right) \left( 2\sigma
_{1}+\sigma _{3}\right) -\rho \left( \sigma _{1}-\sigma _{3}\right) ^{2}]\}
\end{eqnarray*}

\bigskip

\begin{eqnarray*}
C_{1} &=&-2E_{1}^{2}\left[ 3\mathcal{E}_{z}\sigma _{1}+2\gamma \left( \sigma
_{1}-10\sigma _{2}\right) \right] \\
&&+E_{1}\{4E_{2}\left[ 3\mathcal{E}_{z}\sigma _{1}+2\gamma \left( \sigma
_{1}+8\sigma _{2}\right) \right] \\
&&+3\left( \sigma _{2}-\sigma _{1}\right) \left[ \mathcal{E}_{z}\left(
10\sigma _{1}^{2}+4\sigma _{1}\sigma _{2}-5\sigma _{2}^{2}\right) +2\gamma
\left( 14\sigma _{1}^{2}+20\sigma _{1}\sigma _{2}+11\sigma _{2}^{2}\right) %
\right] \} \\
&&-2E_{2}^{2}\left[ 3\mathcal{E}_{z}\sigma _{1}+2\gamma \left( 19\sigma
_{1}+8\sigma _{2}\right) \right] \\
&&-3E_{2}\left( \mathcal{E}_{z}+2\gamma \right) \left( \sigma _{1}-\sigma
_{2}\right) \left( 2\sigma _{1}^{2}-7\sigma _{1}\sigma _{2}-4\sigma
_{2}^{2}\right) \\
&&-\rho \left( \sigma _{2}-\sigma _{1}\right) \left[ 3\mathcal{E}_{z}\sigma
_{1}\left( \sigma _{1}-\sigma _{2}\right) +2\gamma \left( \sigma
_{1}^{2}-11\sigma _{1}\sigma _{2}-8\sigma _{2}^{2}\right) \right] \\
&& \\
C_{2} &=&6\{2E_{1}^{2}-E_{1}\left[ 4E_{2}+3\left( \sigma _{2}-\sigma
_{1}\right) \left( 2\sigma _{1}+\sigma _{2}\right) \right] \\
&&+2E_{2}^{2}+3E_{2}\left( \sigma _{2}-\sigma _{1}\right) \left( 2\sigma
_{2}+\sigma _{1}\right) -\rho \left( \sigma _{1}-\sigma _{2}\right) ^{2}]\}.
\end{eqnarray*}
The above equations correspond to the covariant form of the 1+3 orthonormal
frame expressions (72)-(73) of \cite{Elst-Uggla-Lesame-Ellis-Maartens}.

In the reduction to the Petrov type D models i.e. as one takes the ``limit'' 
$S_{I}\rightarrow S_{D}$ by \emph{successively} substituting $\gamma =0$ and 
$E_{3}=-2E_{2},\sigma _{3}=-2\sigma _{2}$, the new constraints must lead 
\emph{smoothly} to identically satisfied relations. Therefore the
consistency ($\Leftrightarrow $ existence) of SIIS Petrov type I models
implies that, in this ``limit'', the constraints (\ref{spatialderivesofshear}%
) (and also equations (\ref{newXconstraint})-(\ref{newZconstraint})) will be
reduced to the corresponding constraints of the Petrov type D models. Indeed
we can verify that for $S_{I}\rightarrow S_{D}$, equations (\ref
{newXconstraint})-(\ref{newZconstraint}) and (\ref{spatialderivesofshear})
lead to the set (\ref{xshearconstraint})-(\ref{zelectricconstraint}) when we
specialize it to $S_{D}$ models.

On the other hand the vanishing of the polynomials $A,B,C$ in the equations (%
\ref{spatialderivesofshear}) gives relations between the kinematical
variables of the spacelike congruences of the form $\alpha \sim \mathcal{E}%
_{x},\beta \sim \mathcal{E}_{y},\gamma \sim \mathcal{E}_{z}$. At the
``limit'' $S_{I}\rightarrow S_{D}$ these relations reduce to $\alpha =%
\mathcal{E}_{x}/2,\beta =-\mathcal{E}_{y}/2,\gamma =0$. We observe that the
third equation reproduces the Petrov type D property $\gamma =0$. The rest
of the ``limiting'' values suggest that in this case, \emph{if a Petrov type
I model exists it should be reduced (in that ``limit'') to the subclass of
Petrov type D models satisfying $\alpha =\mathcal{E}_{x}/2$ and $\beta =-%
\mathcal{E}_{y}/2$}. In fact we can easily show that this case corresponds
to the Locally Rotationally Symmetric (LRS) models of Ellis class II \cite
{Ellis-Elst1,Elst-Ellis}.

A further propagation of (\ref{spatialderivesofshear}) and the use of (\ref
{sheareigenvaluesevolution})-(\ref{energyconservation2}), (\ref
{evolutionxcongruence})-(\ref{evolutionzcongruence}) and (\ref
{xcommutaionrelation})-(\ref{zcommutaionrelation}) give algebraic relations
of the form:\ 
\begin{equation}
\alpha =\mathcal{E}_{x}f_{1}(E_{2},E_{3},\theta ,\rho ,\sigma _{2},\sigma
_{3})  \label{alpharelation}
\end{equation}
\begin{equation}
\beta =\mathcal{E}_{y}f_{2}(E_{1},E_{3},\theta ,\rho ,\sigma _{1},\sigma
_{3})  \label{betarelation}
\end{equation}
\begin{equation}
\gamma =\mathcal{E}_{z}f_{3}(E_{1},E_{2},\theta ,\rho ,\sigma _{1},\sigma
_{2})  \label{gammarelation}
\end{equation}
where $f_{1},f_{2},f_{3}$ are rational functions in which both the
numerators and denominators are fourth order polynomials with respect to the
factor $E_{2}-E_{1}$ and their corresponding coefficients are fifth order
polynomials in $\sigma _{2}-\sigma _{1}$.

As expected, the propagation of equations (\ref{spatialderivesofshear}) is
also not identically satisfied which means that (\ref{alpharelation})-(\ref
{gammarelation}) represent another set of non-trivial constraints.
Proceeding in a similar way as before, we take the ``limit'' $%
S_{I}\rightarrow S_{D}$ and expect to obtain identities since, at this level
of differentiation, there are no corresponding equations for the Petrov type
D models. However, it can be shown that (with an obvious abuse of
notation):\ 
\begin{equation}
\lim_{S_{I}\rightarrow S_{D}}f_{1}=\frac{1}{2},\hspace{0.2cm}%
\lim_{S_{I}\rightarrow S_{D}}f_{2}=-\frac{1}{2},\hspace{0.2cm}%
\lim_{S_{I}\rightarrow S_{D}}f_{3}=0.  \label{limits1}
\end{equation}
At this stage, the values of the ``limits'' (\ref{limits1}) can be
understood due to the specific structure of the rational functions (\ref
{alpharelation})-(\ref{gammarelation}). In particular, the linear parts of
the polynomial expressions in $f_{1}$ and $f_{2}$ are proportional to each
other with coefficients $1/2$ and $-1/2$ respectively. 
%\footnote{We would like to thank an anonymous referee for pointing this to us.}
The ratio of the associated linear parts in $f_{3}$ scales as $\sim
(E_{3}+2E_{2})(\sigma _{3}+2\sigma _{2}$) which implies that $%
f_{3}\rightarrow 0$ for $S_{I}\rightarrow S_{D}$. Moreover, the
indeterminate cases where either the two pairs of numerators and
denominators or the linear parts of $f_{1},f_{2}$ vanish, lead to algebraic
relations $\theta =\theta (\sigma _{1},...)$. However, it can be easily
verified that none of these solutions is consistent with the evolution
equations (\ref{evolutionexpansion})-(\ref{evolutionshear}) and (\ref
{evolutionelectric})-(\ref{energyconservation}) except if $\Lambda \neq 0$,
which in turn implies the existence of SIIS models of the Petrov type I with
non-vanishing cosmological constant \cite{VandenBergh-Wylleman1}. In the
Appendix, we give the solutions of the indeterminate cases and for
illustration purposes, we demonstrate the inconsistency of one of these
solutions when the cosmological constant vanishes.

Assuming that $\mathcal{E}_{x},\mathcal{E}_{y},\mathcal{E}_{z}\neq 0$, from
the evolution equations (\ref{evolutionxcongruence})-(\ref
{evolutionzcongruence}) and for every Petrov type I model, we obtain the
following equations: 
\begin{equation}
\left( f_{1}\right) ^{\cdot }=\left( \frac{\alpha }{\mathcal{E}_{x}}\right)
^{\cdot }=(1-4f_{1}^{2})(\sigma _{3}-\sigma _{2})/4\Rightarrow \left( \ln
\left| \frac{2f_{1}+1}{2f_{1}-1}\right| \right) ^{\cdot }=\sigma _{3}-\sigma
_{2}  \label{functionsandshear1}
\end{equation}
\begin{equation}
\left( f_{2}\right) ^{\cdot }=\left( \frac{\beta }{\mathcal{E}_{y}}\right)
^{\cdot }=(1-4f_{2}^{2})(\sigma _{1}-\sigma _{3})/4\Rightarrow \left( \ln
\left| \frac{2f_{2}+1}{2f_{2}-1}\right| \right) ^{\cdot }=\sigma _{1}-\sigma
_{3}  \label{functionsandshear2}
\end{equation}
\begin{equation}
\left( f_{3}\right) ^{\cdot }=\left( \frac{\gamma }{\mathcal{E}_{z}}\right)
^{\cdot }=(1-4f_{3}^{2})(\sigma _{2}-\sigma _{1})/4\Rightarrow \left( \ln
\left| \frac{2f_{3}+1}{2f_{3}-1}\right| \right) ^{\cdot }=\sigma _{2}-\sigma
_{1}.  \label{functionsandshear3}
\end{equation}
It should be noted that in the last equations, we have calculated the time derivatives of the functions $f_\alpha$ by using 
the evolution equations (\ref{evolutionxcongruence})-(\ref{evolutionzcongruence}). In this sense, the initial constraints have been time propagated only \emph{up to third order},  
avoiding to compute the derivatives directly from the highly complicated expressions (\ref{alpharelation})-(\ref{gammarelation}).  
  
Expressing equations (\ref{functionsandshear1})-(\ref{functionsandshear3})
in (local) coordinate form, a straightforward calculation yields: 
\begin{equation}
f_{1}=\frac{e^{\Gamma +D_{1}(x,y,z)}\mp e^{B}}{2(e^{\Gamma +D_{1}(x,y,z)}\pm e^{B})%
}  \label{functions1}
\end{equation}
\begin{equation}
f_{2}=\frac{e^{A+D_{2}(x,y,z)}\mp e^{\Gamma }}{2(e^{A+D_{2}(x,y,z)}\pm e^{\Gamma })%
}  \label{functions2}
\end{equation}
\begin{equation}
f_{3}=\frac{e^{B+D_{3}(x,y,z)}\mp e^{A}}{2(e^{B+D_{3}(x,y,z)}\pm e^{A})}
\label{functions3}
\end{equation}
where $D_{\alpha }(x,y,z)$ are smooth functions of the spatial coordinates.

Equations (\ref{functions1})-(\ref{functions3}) imply that for $%
S_{I}\rightarrow S_{D}\Rightarrow A=B+H$ we have $f_{3}=0\Rightarrow
D_{3}(x,y,z)=H(x,y)$. However $f_{1}\neq 1/2$ and $f_{2}\neq -1/2$ for \emph{%
any} smooth functions $A,B,D_{1},D_{2}$. This contradiction shows that the $%
x,y$ expansion rates must vanish $\mathcal{E}_{x}=\mathcal{E}_{y}=0$ which
by means of equations (\ref{evolutionxcongruence})-(\ref
{evolutionycongruence}) implies that $\alpha =\beta =0$. Hence, due to the
Killing conditions (\ref{ConformalCondition3}), there exists a (necessarily)
Abelian group of isometries acting on 2-dimensional spacelike orbits.
Furthermore, equations (\ref{xproperties}) and (\ref{yproperties}) imply
that each of the spacelike KVFs is hypersurface orthogonal, therefore the
SIIS models of Petrov type I are reduced to the subclass of diagonal $G_{2}$
models \cite{vanElst:2001xm}. Using a similar procedure we can further show
that $\mathcal{E}_{z}=\gamma =0$. Consequently, from equations (\ref
{ConformalCondition3}) or, equivalently, the correspondence (\ref
{n-shearrelations})-(\ref{a-expansionrelations}), it follows that the Petrov
type I model \emph{always} admits three independent and commuting spacelike
KVFs and the resulting silent model being SH of the Bianchi type I.

It should be remarked that the above methodology cannot be applied when: a)
two\footnote{%
Using the Ricci identities (\ref{spatialXRicciIdentity})-(\ref
{spatialZRicciIdentity}) for the spatial triad $\left\{
x^{a},y^{a},z^{a}\right\} $ it can be shown that e.g. $\alpha =\mathcal{E}%
_{x}/2$ implies $\mathcal{E}_{x}=0=\alpha $ or $\beta =-\mathcal{E}_{y}/2$.}
of the functions $f_{\alpha }$ are equal to $\pm 1/2$ or b) when $%
f_{1}=f_{2}=0$ and $f_{3}=\pm 1/2$ (a special class of the diagonal $G_{2}$
family of perfect models). Nonetheless, both cases are incompatible with the
full set of the dynamical equations. In particular, in the diagonal $G_{2}$
class of models, satisfying $f_{3}=\pm 1/2\Leftrightarrow \gamma =\pm 
\mathcal{E}_{z}/2$, equations (\ref{xpropagationequationexpansion})-(\ref
{zpropagationequationshear}) imply that 
\begin{equation}
E_{1}=-\frac{\theta ^{2}-3\theta \sigma _{1}-3\left( \rho +3\sigma
_{1}\sigma _{2}+3\sigma _{2}^{2}\right) }{9},\hspace{0.2cm}E_{2}=\frac{%
2\theta ^{2}+3\theta \sigma _{2}-6\rho -9\sigma _{2}^{2}}{9}.
\label{electriceigenvalues}
\end{equation}
However, from (\ref{zshearconstraint}), (\ref{zelectricconstraint}) and (\ref
{zdivergence}) it follows that the quantities $w_{1}=\sigma _{2}+\theta /3$
and $w_{2}=E_{2}+\rho /6$ are \emph{spatially homogeneous}. Using the
commutation relation (\ref{zcommutaionrelation}), we find that the algebraic
equation $\left( \dot{w}_{2}\right) ^{\symbol{126}}=0$ has as a solution $%
\sigma _{2}=\sigma _{2}(w_{1},w_{2},\sigma _{1})$. Substituting in the
initial relation $f_{3}=\pm 1/2$ we finally compute $\sigma _{1}=\sigma
_{1}(w_{1},w_{2})$ and $\sigma _{2}=\sigma _{2}(w_{1},w_{2})$ i.e. the shear
eigenvalues are also spatial homogeneous which in turn implies that $%
\mathcal{E}_{z}=0=\gamma $. Using a similar method we can easily show that
the conditions $f_{1}=1/2 $ and $f_{2}=-1/2$ lead also to $\mathcal{E}_{x}=%
\mathcal{E}_{y}=0=\alpha=\beta$ and should be excluded.

We conclude this section by noticing that the above ``limiting process'' is
always possible within the SIIS solutions of the EFE due to their specific
geometrical and dynamical behaviour. In particular, let us assume that a
Petrov type I model exists and is completely disconnected (in the spirit of
the above reduction) from the Petrov type D models (for example, due to the
occurrence of denominators of the form $(E_{2}-E_{1})$ etc.). Then, this
must be true at any point of the spacetime manifold or, strictly speaking,
during the time evolution of the model. Nevertheless, it has been shown that
(apart from a set of measure zero) for initial states within the Petrov type
I invariant set, the orbits approach at the asymptotic regimes (i.e. at
early or late times), either a Petrov type D model or a vacuum SH model of
Bianchi type I \cite
{Bruni-Matarrese-Pantano1,Bruni-Matarrese-Pantano2,Wainwright-Ellis-Book}.
The exceptional initial states are treated similarly. From equations (\ref
{functionsandshear1})-(\ref{functionsandshear3}) it turns out that $%
f_{\alpha }$ are monotone functions of the time coordinate $t$ or the
dimensionless time variable $\tau $ which is defined according to $dt/d\tau
=3/\theta $. In addition and taking into account the relations (\ref
{functionsandshear1})-(\ref{functionsandshear3}), the functions $f_{\alpha }$ satisfy 
$f_{\alpha}\rightarrow \pm 1/2$ as $\tau \rightarrow \pm \infty $ which due to
equations (\ref{spatialXRicciIdentity})-(\ref{spatialZRicciIdentity}) and
the previous discussion implies that $\mathcal{E}_{x}=\mathcal{E}_{y}=%
\mathcal{E}_{z}=0$ and the model will approach a SH model of Bianchi type I. 
%This shows that the reduction $S_{I}\rightarrow S_{D}$ can always be made at least within the family of SIIS models.

\section{Discussion}

In the present article we have undertaken a detailed covariant analysis of
the question regarding the broadness of vacuum and non-vacuum SIIS models
which is equivalent to the consistency of the spatial div and curl equations
with the evolution equations. By exploiting the basic tools from the theory
of the spacelike congruences and the fact that the shear $\sigma_{ab}$ and
the electric part $E_{ab}$ share a common spatial eigenframe, we have
expressed the constraints (\ref{xshearconstraint})-(\ref{zelectricconstraint}%
) in terms of the spacelike irreducible kinematical variables. These
expressions make clear the significant role that the spacelike expansions
and shear magnitudes play in the sense that they influence the overall
spatial variation of the kinematical and dynamical variables (and vice
versa) of the corresponding models. Therefore in order to have a transparent
and complete picture of the dynamics of SIIS models, we have incorporated
the associated time and spatial propagation equations (\ref
{evolutionxcongruence})-(\ref{zpropagationequationshear}) of the spacelike
kinematical quantities.

Obviously, the suggested procedure has strong similarities with the 1+3 ONF
approach and the basic steps in analyzing the SIIS are essentially the same,
which shows the covariant nature of the incompatibility of the constraints (%
\ref{xshearconstraint})-(\ref{zelectricconstraint}) with the evolution
equations. However, the approach followed here using elements from the
theory of spacelike congruences, has allowed us to interpret the
implications of the constraints (\ref{xshearconstraint})-(\ref
{zelectricconstraint}) in a geometrical manner by expressing them as a set
of conditions that must be satisfied by the kinematical quantities of the
family of the spacelike curves.

It is worth noticing that, since the spacelike vector fields have been
chosen in a \emph{unique} way (as eigenvectors of the shear tensor and the
electric part of the Weyl tensor), these conditions represent \emph{invariant%
} relations. This fact facilitated the analytical study of the induced
constraints in a (local) coordinate system and allowed us to show that the
SH models of Bianchi type I are the only compatible Petrov type I spacetimes
within the family of silent models.

We note that, the present approach permits us to draw further conclusions
for more general spatially inhomogeneous configurations of the Petrov type
I. For example, even when we allow the presence of a fluid with non-zero
pressure (losing in that way the ``silence'' property), Petrov type D
models are still the only permissible. This can be justified by observing
that the pressure is necessary spatially homogeneous \cite
{Spero-Szafron} (because $\dot{x}^{a}=\dot{y}^{a}=%
\dot{z}^{a}=0$), hence when we apply the commutation relations (\ref
{xcommutaionrelation})-(\ref{zcommutaionrelation}) in the evolution of the
eigenvalues of the electric part $E_{ab}$ (into which the fluid pressure
enters) the form of the new constraints (\ref{newXconstraint})-(\ref
{newZconstraint}) is not affected. In addition, the fact that we didn't use
equation (\ref{evolutionexpansion}), shows that the uniqueness of Petrov
type D models is also \emph{independent}, in general, from the presence of a
non-zero cosmological constant. Nonetheless, an exception to this conclusion
is when the eigenvalues of the shear tensor $\sigma _{ab}$ are constrained
to be proportional to $\theta $, due to the appearance of indeterminate
expressions. In this case, a non-zero pressure or the cosmological constant
play a key role in the existence of a broader class of SIIS solutions apart
from the SH models of Bianchi type I \cite{VandenBergh-Wylleman1}. Finally
we expect that this approach can be used in order to check how large is the
family of SIIS models by relaxing the vanishing of the vorticity and the
magnetic part of the Weyl tensor but maintaining their ``silence''
properties i.e. the vanishing of the curls of $E_{ab},H_{ab}$ and the
pressure $p$. %\clearpage

\section*{Appendix}

\setcounter{equation}{0}

In this Appendix, we present the solutions of the indeterminate cases i.e.
when either the two pairs of numerators and denominators or the linear parts
of $f_{1},f_{2}$ vanish. In particular, solving the algebraic equations num($%
f_{1,2}$)$=0$ and denom($f_{1,2}$)$=0$ we obtain\footnote{%
We note that we have ignored the solutions $\sigma _{1}=0$ and $\sigma
_{1}+\sigma _{2}=0$ since they lead to unphysical models ($\rho <0$).}: 
\begin{eqnarray}
\sigma _{2} &=&-\frac{\sigma _{1}(4E_{1}+5E_{2})}{5E_{1}+4E_{2}},\hspace{%
0.2cm}\theta =\frac{3\sigma _{1}\left( E_{1}-E_{2}\right) }{5E_{1}+4E_{2}}, 
\nonumber \\
&&  \nonumber \\
\rho &=&\frac{2\left[ 25E_{1}^{2}+E_{1}\left( 40E_{2}+27\sigma
_{1}^{2}\right) +E_{2}\left( 16E_{2}+27\sigma _{1}^{2}\right) \right] }{%
9\sigma _{1}^{2}}  \label{solution1}
\end{eqnarray}
\vskip 0.5cm 
\begin{equation}
\sigma _{2}=\frac{\sigma _{1}(2E_{1}+E_{2})}{4E_{2}-E_{1}},\hspace{0.2cm}%
\theta =\frac{3\sigma _{1}\left( 2E_{1}+E_{2}\right) }{E_{1}-4E_{2}},\hspace{%
0.2cm}\rho =\frac{2\left[ E_{1}^{2}-8E_{1}E_{2}+E_{2}\left( 16E_{2}-27\sigma
_{1}^{2}\right) \right] }{9\sigma _{1}^{2}}  \label{solution2}
\end{equation}
\vskip 0.5cm 
\begin{equation}
\sigma _{2}=\frac{\sigma _{1}(4E_{1}-E_{2})}{E_{1}+2E_{2}},\hspace{0.2cm}%
\theta =-3\sigma _{1},\hspace{0.2cm}\rho =\frac{2\left[ E_{1}^{2}+E_{1}%
\left( 4E_{2}-27\sigma _{1}^{2}\right) +4E_{2}^{2}\right] }{9\sigma _{1}^{2}}%
.  \label{solution3}
\end{equation}
On the other hand, the vanishing of the linear parts of $f_{1},f_{2}$ give
the following solution: 
\begin{equation}
\theta =-3\sigma _{2},\hspace{0.2cm}\rho =\frac{2E_{2}\left( E_{2}-3\sigma
_{2}^{2}\right) }{\sigma _{2}^{2}}  \label{solution4}
\end{equation}
As we have mentioned in Section 3, none of the above solutions are
consistent with the evolution equations when $\Lambda =0$. For illustration
purposes, we show the inconsistency of the solution (\ref{solution2}).
Propagating the second relation of (\ref{solution2}) and using the evolution
equations (\ref{evolutionexpansion}) and (\ref{sheareigenvaluesevolution})-(%
\ref{energyconservation2}) we get the algebraic equation: 
\[
E_{1}^{2}-8E_{1}E_{2}+16E_{2}^{2}-9\Lambda \sigma _{1}^{2}=0 
\]
with solutions $E_{1}=4E_{2}\pm 3\sqrt{\Lambda }\sigma _{1}$. Obviously,
these solutions are valid only for non-zero (and positive) cosmological
constant. Similarly, we can easily show the inconsistency for the rest
solutions for the case $\Lambda =0$. \vskip1.0cm

\noindent \textbf{Acknowledgments}\newline
We would like to thank H. van Elst for useful comments and remarks. 
The authors gratefully acknowledge financial support from the Spanish
Ministerio de Educaci\'{o}n {y} Ciencia through research grants SB2004-0110
(PSA) and FPA2004-03666 (JC). \vskip 1.0cm %\clearpage


\begin{thebibliography}{99}
\bibitem{Szekeres1}  Szekeres P, \emph{A class of inhomogeneous cosmological
models}, 1975 Comm. Math. Phys. \textbf{41} 55-64.

\bibitem{Bonnor-Tomimura1}  Bonnor W B and Tomimura N, \emph{Evolution of
Szekeres's cosmological models}, 1976 Mon. Not. R. Astron. Soc. \textbf{175}
85-93.

\bibitem{Szafron}  Szafron D A, \emph{Inhomogeneous cosmologies: New exact
solutions and their evolution}, 1977 J. Math. Phys. \textbf{18} 1673-1677.

\bibitem{Wainwright}  Wainwright J, \emph{Characterization of the Szekeres
inhomogeneous cosmologies as algebraically special space-times}, 1977 J.
Math. Phys. \textbf{18} 672-675.

\bibitem{Spero-Szafron}  Spero A and Szafron D A, \emph{Spatial conformal
flatness in homogeneous and inhomogeneous cosmologies}, 1978 J. Math. Phys. 
\textbf{19} 1536-1541.

\bibitem{Goode-Wainwright}  Goode S W and Wainwright J, \emph{Singularities
and evolution of the Szekeres cosmological models}, 1982 Phys. Rev. D 
\textbf{26} 3315-3326.

\bibitem{Bertschinger-Jain}  Bertschinger E and Jain B, \emph{Gravitational
instability of cold matter}, 1994 Astrophys.J. \textbf{431} 486-494 (\emph{%
Preprint} astro-ph/9307033).

\bibitem{Barnes-Rowlingson}  Barnes A and Rowlingson R R, \emph{Irrotational
perfect fluids with a purely electric Weyl tensor}, 1989 Class. Quantum
Grav. \textbf{6} 949-960.

\bibitem{Krasinski}  Krasi\'{n}ski A, \emph{Inhomogeneous cosmological models%
}, (Cambridge University Press, Cambridge 1997).

\bibitem{Goode}  Goode S W, \emph{Analysis of spatially inhomogeneous
perturbations of the FRW cosmologies}, 1989 Phys. Rev. D \textbf{39}
2882-2892.

\bibitem{Matarrese-Pantano-Saez1}  Matarrese S, Pantano O and Saez D, \emph{%
General relativistic approach to the nonlinear evolution of collisionless
matter}, 1993 Phys. Rev. D \textbf{47} 1311-1323.

\bibitem{Matarrese-Pantano-Saez2}  Matarrese S, Pantano O and Saez D, \emph{%
General relativistic dynamics of irrotational dust:\ cosmological
implications}, 1994 Phys. Rev. Lett. \textbf{72} 320-323 (\emph{Preprint}
astro-ph/9310036).

\bibitem{Matarrese-Pantano-Saez3}  Matarrese S, Pantano O and Saez D, \emph{%
A relativistic approach to gravitational instability in the expanding
universe: second order Lagrangian perturbations}, 1994 Mon. Not. R. Astron.
Soc. \textbf{271} 513-522 (\emph{Preprint} astro-ph/9403032).

\bibitem{Kofman-Pogosyan}  Kofman L and Pogosyan D, \emph{Dynamics of
gravitational instability}, 1995 Astrophys. J. \textbf{442} 30-38 (\emph{%
Preprint} astro-ph/9403029).

\bibitem{Bruni-Matarrese-Pantano1}  Bruni M, Matarrese S and Pantano O, 
\emph{Dynamics of silent universes}, 1995 Astrophys. J. \textbf{445} 958-977
(\emph{Preprint} astro-ph/9406068).

\bibitem{Bruni-Matarrese-Pantano2}  Bruni M, Matarrese S and Pantano O, 
\emph{A local view of the observable universe}, 1995 Phys. Rev. Lett. 
\textbf{74} 1916-1919 (\emph{Preprint} astro-ph/9407054).

\bibitem{Lesame-Dunsby-Ellis}  Lesame W M, Dunsby P K S and Ellis G F R, 
\emph{Integrability conditions for irrotational dust with a purely electric
Weyl tensor: a tetrad analysis}, 1995 Phys. Rev. D \textbf{52} 3406-3415 (%
\emph{Preprint} astro-ph/9410005).

\bibitem{Lesame-Ellis-Dunsby}  L Lesame W M, Ellis F G R, Dunsby P K S, 
\emph{Irrotational dust with Div H=0}, 1996 Phys. Rev. D \textbf{53} 738-746
(\emph{Preprint} gr-qc/9508049).

\bibitem{Bonilla-Mars-Senovilla-Sopuerta-Vera}  Bonilla M A G, Mars M,
Senovilla J M M, Sopuerta C F and Vera R, \emph{Comment on: ``Integrability
conditions for irrotational dust with a purely electric Weyl tensor: a
tetrad analysis''}, 1996 Phys. Rev. D \textbf{54} 6565-6566.

\bibitem{ElstPhD}  van Elst H, \emph{Extensions and applications of 1+3
decomposition methods in general relativistic cosmological modelling}, 1996
Ph. D. Thesis, Queen Mary and Westfield College, London.

\bibitem{Maartens-Waves}  Maartens R, \emph{Linearization instability of
gravity waves?}, 1997 Phys. Rev. D \textbf{55} 463-467 (\emph{Preprint}
astro-ph/9609198).

\bibitem{Elst-Uggla-Lesame-Ellis-Maartens}  van Elst H, Uggla C, Lesame W M,
Ellis G F R and Maartens R, \emph{Integrability of irrotational silent
cosmological models}, 1997 Class. Quantum Grav. \textbf{14} 1151-1162 (\emph{%
Preprint} gr-qc/9611002).

\bibitem{Sopuerta}  Sopuerta C F, ``\emph{New study of silent universes}'',
1997 Phys. Rev. D \textbf{55}, 5936-5950.

\bibitem{Mars}  Mars M, ``\emph{3 + 1 description of silent universes: a
uniqueness result for the Petrov type I vacuum case}'', 1999 Class. Quantum
Grav. \textbf{16} 3245-3262 (\emph{Preprint} gr-qc/9909089).

\bibitem{Wainwright-Ellis-Book}  Wainwright J and Ellis G F R (Editors), 
\emph{Dynamical Systems in Cosmology }(Cambridge University Press, Cambridge
1997).

\bibitem{Carot-Sintes}  Carot J and Sintes A M, \emph{Homothetic perfect
fluid spacetimes}, 1997 Class. Quantum Grav. \textbf{14} 1183-1205 (\emph{%
Preprint} gr-qc/9607061).

\bibitem{Carr-Coley1}  Carr B J and Coley A A, \emph{Self-similarity in
general relativity}, 1999 Class. Quantum Grav. \textbf{16} R31-R71 (\emph{%
Preprint} gr-qc/9806048).

\bibitem{Barrow-Hervik1}  Barrow J D and Hervik S, \emph{The future of
tilted Bianchi universes}, 2003 Class. Quantum Grav. \textbf{20} 2841-2854 (%
\emph{Preprint} gr-qc/0304050).

\bibitem{Apostol1}  Apostolopoulos P S, \emph{Self-similar Bianchi models:
I. Class A models}, 2003 Class. Quantum Grav. \textbf{20} 3371-3384 (\emph{%
Preprint} gr-qc/0306119).

\bibitem{Apostol2}  Apostolopoulos P S, \emph{On tilted perfect fluid
Bianchi type VI}$_{0}$\emph{\ self-similar models}, 2004 Gen. Relativ. Grav. 
\textbf{36} 1939-1945 (\emph{Preprint} gr-qc/0310033).

\bibitem{Apostol3}  Apostolopoulos P S, \emph{Self-similar Bianchi models:
II. Class A models}, 2005 Class. Quantum Grav. \textbf{22} 323-338 (\emph{%
Preprint} gr-qc/0411102).

\bibitem{Hervik1}  Hervik S, \emph{The asymptotic behaviour of tilted
Bianchi type VI}$_{0}$\emph{\ universes}, 2004 Class. Quantum Grav. \textbf{%
21} 4425-4441 (\emph{Preprint} gr-qc/0403040).

\bibitem{Coley-Hervik1}  Coley A A and Hervik S, \emph{Bianchi cosmologies:
A tale of two tilted fluids}, 2004 Class. Quantum Grav. \textbf{21}
4193-4208 (\emph{Preprint} gr-qc/0406120).

\bibitem{Apostol4}  Apostolopoulos P S, \emph{Equilibrium points of the
tilted perfect fluid Bianchi type VI}$_{h}$ \emph{state space}, 2005 Gen.
Relativ. Grav. \textbf{37} 937-952 (\emph{Preprint} gr-qc/0407040).

\bibitem{Hervik-Hoogen-Coley}  Hervik S, van den Hoogen R and Coley A A, 
\emph{Future asymptotic behaviour of tilted Bianchi models of type IV and VII%
}$_{h}$, 2005 Class. Quantum Grav. \textbf{22} 607-633 (\emph{Preprint}
gr-qc/0409106).

\bibitem{Coley-Hervik2}  Coley A A and Hervik S, \emph{A dynamical systems
approach to the tilted Bianchi models of solvable type}, 2005 Class. Quantum
Grav. \textbf{22} 579-605 (\emph{Preprint} gr-qc/0409100).

\bibitem{Hervik-Hoogen-Lim-Coley}  Hervik S, van den Hoogen R J, Lim W C,
Coley A A, \emph{The futures of Bianchi type VII$_{0}$ cosmologies with
vorticity}, 2005 Class. Quantum Grav. \textbf{23} 845-866 (\emph{Preprint }%
gr-qc/0509032).

\bibitem{Carr-Coley2}  Carr B J and Coley A A, \emph{The Similarity
Hypothesis in General Relativity}, 2005 Gen. Relativ. Grav. \textbf{37}
2165-2188 (\emph{Preprint} gr-qc/0508039).

\bibitem{vanElst:2001xm}  van Elst H, Uggla C and Wainwright J, \emph{%
Dynamical systems approach to $G_{2}$ cosmology}, 2002 Class.\ Quantum Grav. 
\textbf{19} 51-82 (\emph{Preprint} gr-qc/0107041). 
%%CITATION = GR-QC 0107041;%%

\bibitem{VandenBergh-Wylleman2}  Wylleman L and van den Bergh N, ``\emph{%
Petrov type I silent universes with G}$_{3}$\emph{\ isometry group: the
uniqueness result recovered}'', 2006 Class. Quantum Grav. \textbf{23}
329-334 (\emph{Preprint} gr-qc/0508092).

\bibitem{Barrow-Stein-Schabes}  Barrow J and Stein-Schabes J, \emph{%
Inhomogeneous cosmologies with cosmological constant}, 1984 Phys. Lett. A. 
\textbf{103} 315-317.

\bibitem{VandenBergh-Wylleman1}  van den Bergh N and Wylleman L, \emph{%
Silent universes with a cosmological constant}, 2004 Class. Quantum Grav. 
\textbf{21} 2291-2299 (\emph{Preprint} gr-qc/0402125).

\bibitem{Greenberg1}  Greenberg P J, \emph{The general theory of spacelike
congruences with an application to vorticity in relativistic hydrodynamics},
1970 J. Math. Anal. Appl. \textbf{30} 128-143.

\bibitem{Tsamparlis-Mason}  Tsamparlis M and Mason D P, \emph{On spacelike
congruences in general relativity}, 1983 J. Math. Phys. \textbf{24}
1577-1593.

\bibitem{Mason-Tsamparlis}  Mason D P and Tsamparlis M, \emph{Spacelike
conformal Killing vectors and spacelike congruences}, 1985 J. Math. Phys. 
\textbf{27} 2881-2901.

\bibitem{Saridakis-Tsamparlis}  Saridakis E and Tsamparlis M, \emph{Symmetry
inheritance of conformal Killing vectors}, 1991 J. Math. Phys. \textbf{32}
1541-1551.

\bibitem{Ellis-Elst1}  Ellis G F R and van Elst H, \emph{Cosmological Models}%
, in Carg\`{e}se Lectures 1998, \emph{Theoretical and Observational Cosmology%
} ed M. Lachi\`{e}ze-Rey, Dordrecht: Kluwer (\emph{Preprint }gr-qc/9812046).

\bibitem{Tsamp-Apostol4}  Tsamparlis M and Apostolopoulos P S, \emph{%
Symmetries of Bianchi I space-times}, 2000 J. Math. Phys. \textbf{41}
7573-7588 (\emph{Preprint }gr-qc/0006083).

\bibitem{Elst-Uggla}  van Elst H and Uggla C, \emph{General relativistic 1+3
orthonormal frame approach}, 1997 Class. Quantum Grav. \textbf{14} 2673-2695
(\emph{Preprint} gr-qc/9603026).

\bibitem{Apostol-Tetrad-Appr-SH-Models}  Apostolopoulos P S, \emph{A
geometric description of the intermediate behaviour for spatially
homogeneous models}, 2005 Class. Quantum Grav. \textbf{22} 4425-4442 (\emph{%
Preprint} gr-qc/0506114).

\bibitem{Elst-Ellis}  van Elst H and Ellis G F R, \emph{The covariant
approach to LRS perfect fluid spacetime geometries}, 1996 Class. Quantum
Grav. \textbf{13} 1099-1127 (\emph{Preprint} gr-qc/9510044).

%\cite{vanElst:2001xm}
\end{thebibliography}
\end{document}